\documentclass[aps,prb,preprint,superscriptaddress,floatfix]{revtex4}

\usepackage{graphicx}
\usepackage{dcolumn}
\usepackage{bm}
\usepackage{amsfonts}
\usepackage{amsmath, amsthm, amssymb}
\DeclareGraphicsExtensions{.png}

\newcommand{\eqz}[1]{Eq.~(#1)}
\newcommand{\Fig}[1]{Fig.~{#1}}
\newcommand{\Sch}{Schr\"odinger}
\newcommand{\ky}{k_{2}}
\newcommand{\im}   {\mathrm{i}}		
\newcommand{\ees}   {\mathrm{e}}
\newcommand{\de}{\mathrm{d}}

\begin{document}

\title{Cylindrical Two-Dimensional Electron Gas in a Transverse Magnetic Field}

\author{Giulio Ferrari}
\email[]{giulio.ferrari@unimore.it}
\affiliation{CNR-INFM Research Center on nanoStructures and bioSystems at Surfaces (S3), Via Campi 213/A, 41100 Modena, Italy.}
\author{Andrea Bertoni}
\affiliation{CNR-INFM Research Center on nanoStructures and bioSystems at Surfaces (S3), Via Campi 213/A, 41100 Modena, Italy.}
\author{Guido Goldoni}
\affiliation{CNR-INFM Research Center on nanoStructures and bioSystems at Surfaces (S3), Via Campi 213/A, 41100 Modena, Italy.}
\affiliation{Dipartimento di Fisica, Universit\`a di Modena e Reggio Emilia, Via Campi 213/A, 41100 Modena, Italy.}
\author{Elisa Molinari}
\affiliation{CNR-INFM Research Center on nanoStructures and bioSystems at Surfaces (S3), Via Campi 213/A, 41100 Modena, Italy.}
\affiliation{Dipartimento di Fisica, Universit\`a di Modena e Reggio Emilia, Via Campi 213/A, 41100 Modena, Italy.}

\date{\today}

\begin{abstract}
We compute the single-particle states of a two-dimensional electron gas confined to the surface of a cylinder immersed in a magnetic field.
The envelope-function equation has been solved exactly for both an homogeneous and a periodically modulated magnetic field  perpendicular to the cylinder axis. The nature and energy dispersion of the quantum states reflects the interplay between different lengthscales, namely, the cylinder diameter, the magnetic length, and, possibly, the wavelength of the field modulation.
We show that a transverse homogeneous magnetic field drives carrier states from a quasi-2D (cylindrical) regime to a quasi-1D regime where carriers form channels along the cylinder surface.
Furthermore, a magnetic field which is periodically modulated along the cylinder axis may confine the carriers to tunnel-coupled stripes, rings or dots on the cylinder surface, depending on the ratio between the the field periodicity and the cylinder radius.
Results in different regimes are traced to either incipient Landau levels formation or Aharonov-Bohm behaviour.
\end{abstract}


\maketitle

\section{Introduction\label{sec:intro}}

The interest in the electronic properties of quantum
systems with cylindrical symmetry has received a boost since the early proposals of adopting
carbon nanotubes\cite{Radushkevich52,Oberlin76} as building blocks for
future nanoelectronic devices, exploiting their peculiar mechanical and
electrical properties\cite{Iijima91,Dresselhaus00}.
In recent years new inorganic semiconductor systems are also emerging where
carriers are confined on a bent surface, and several possibilities arise to
obtain 2D electron gases (2DEGs) with cylindrical symmetry (C2DEGs), which may enrich
the wealth of physics and applications of planar semiconductor nanostructures.
One such system can be obtained from a standard epitaxially grown
2DEG at a planar heterojunction, which is
then overgrown with a lattice mismatched material at some distance
above the buried 2DEG. A sacrificial layer below the 2DEG allows the
release of the elastic energy by lift-off and bending of a thin layer
of material embedding the 2DEG, up to complete rolling
\cite{Prinz00,Schmidt01,Lorke03,Shaji07}.  The rolled-up layers stick
together, thus forming a
C2DEG with a radius ranging from tens of nanometers up to
several microns, showing peculiar
magneto-resistance with respect to the corresponding planar
structures \cite{Shaji07}.
Alternatively, a C2DEG can be obtained in coaxial structures which can be fabricated similarly to
standard layered heterostructures, but using a cylindrical substrate rather than the usual planar substrate, for multilayer overgrowth of lattice matched materials. The cylindrical
substrate, in turn, can be obtained by a self-standing single-crystal semiconductor nanowire,
fabricated by seeded growth, either assisted by Au \cite{Westwater98,Martensson03} or Ga \cite{Morral08} nanoparticles, the latter possibility being particularly
promising for high-mobility nano-devices to avoid Au-induced deep level traps.
The resulting C2DEGs will have a diameter determined by the diameter of the nanowire used as a substrate, in the few tens of nm range.

Although carbon nanotubes and C2DEGs share the cylindrical
symmetry of the electronic states, they have very different curvatures.
Carbon nanotubes have typical diameters in the
few nanometers range and are basically quasi-1D systems, while the
diameter of C2DEGs is at least one order of magnitude larger. For
this reason, we expect the electronic properties of the latter systems to be dominated
by size quantisation, as in usual planar heterostructures, rather
than by the atomistic details. In addition,
the effect of an external magnetic field is stronger in C2DEGs than in
carbon nanotubes since the magnetic length of typical fields
is comparable to the lateral dimension, while carbon nanotubes diameters
are much smaller. Therefore, the interplay between the cylinder diameter
and the magnetic length adds a new degree of freedom to manipulate
the quantum states of the carriers.

One may also notice that the field itself may be modulated on the scale of the nanostructure diameter \cite{Ye95,Ye96}. Modulated fields are obtained by deposition of ferromagnetic strips or dots on top of a 2DEG, which may give rise to effective periodic potentials and hence to oscillatory behaviours in the magneto-resistance of planar structures \cite{Yoshioka87,Yagi93}. This is the counterpart of homogeneous magnetic fields applied to modulated periodic 2D structures, which may induce complex and fascinating electronic properties, such as the Hofstadter butterfly \cite{Hofstadter76,Gerhardts89}. It is therefore interesting to couple modulated magnetic fields with structures which have a built-in periodicity of the carrier states on the same length scale, such as the C2DEGs introduced above.

In this paper, we investigate the effect of a transverse magnetic field on the single-particle properties of carriers in a C2DEG. Electronic band structures and density of states are obtained with the magnetic field either uniform or periodically modulated along the cylinder. In the former case, the field does not break the translational invariance along the axis of the cylinder, and the electronic states scale with the ratio between the cylinder diameter and the magnetic length. The system is found to show a transition from a quasi-2D to a quasi-1D behaviour as a function of the field strength. On the contrary, a modulated magnetic field breaks the continuous translational symmetry of the cylindrical 2DEG leading to the formation of energy subbands and gaps opening along the cylinder axis. In such a system the effect of the field depends on the interplay between its intensity (or magnetic length), the radius of the tube and the wavelength of the field modulation. We find that the tailoring of different length scales may lead to the formation of electronic states similar to those observed in arrays of anti-dots \cite{Ensslin90}, with the magnetic field giving rise to a series of regions where the carriers can not penetrate, if not belonging to very high energy bands. Allowed regions form a network of tunnel-coupled dots, rings or arrays, depending on the ratio between the cylinder radius and the field periodicity.

The paper is organised as follows.
In Sec.\  \ref{sec:cylinder}, the cylindrical system under study is introduced and its \Sch\ equation is derived.
In Sec.\  \ref{sec:constant}, the theory is applied to the case of an homogeneous transverse field and different regimes of 1D localisation are discussed.
Section \ref{sec:variable} addresses the effect of a spatially-modulated magnetic field in different  regimes of tube radius, field modulation wavelength and field strength. We also discuss our results in connection with Aharonov-Bohm effects and Landau levels formation.
In the last section, the conclusions are drawn.

\section{Hamiltonian of a C2DEG in a magnetic field\label{sec:cylinder}}

We consider the problem of a spinless electron bound to the surface $S$ of a cylinder in a magnetic field perpendicular to its axis.
The general derivation of a proper quantum equation of motion on a
curved surface is a problem with a long history. The classical approaches are the Lagrangian method of de Witt \cite{DeWitt57} and the limiting procedure of Jensen-Koppe  \cite{Jensen71} and da Costa \cite{daCosta81}.
While the former includes the effects of the curvature directly in the equation of motion, the latter is most appropriate when the system under consideration is a real surface embedded in the 3D space \cite{Encinosa03,Marchi05,Meyer07,Taira07}, as in our case.
When a magnetic field is introduced, the problem is much more complicated, and only recently it has been shown, by one of the authors\cite{Ferrari08}, that an analytical expression of the \Sch\ equation where the dynamics on the surface is decoupled from the transverse one can always be obtained provided a proper choice of the gauge is made.
However, for the present case of a straight cylinder with constant radius $r$ the solution is not difficult.

Let the cylinder axis be along the $y$ direction of a Cartesian reference system, with $x$ and $z$ perpendicular to $S$ (see \Fig{\ref{fig:homotube}}). We also define two coordinates $q_1$ and $q_2$ lying on $S$, and $q_3$ perpendicular to $S$, as shown in \Fig{\ref{fig:homotube}}. Therefore, $S$ is defined by
\begin{eqnarray}
\left\{
\begin{array}{l}
x=r\sin(q_1/r),\\
y=q_2,\\
z=r\cos(q_1/r),
\end{array}
\right.
&
\textrm{with}
&
\left\{
\begin{array}{l}
q_1\in[0;2\pi),\\
q_2\in(-\infty;\infty),\\
q_3=0.
\end{array}
\right.
\end{eqnarray}
Next, we consider a magnetic field perpendicular to the cylinder axis. For definiteness, we set the magnetic field along the $z$ axis,
$\vec{B}=(0;0;B(y))$, with an intensity $B(y)$ possibly modulated along the $y$ direction.
A convenient choice for the vector potential is $\vec{A}=(0;xB(y);0)$, which in the cylindrical frame is
\begin{equation}
\label{eq:genvecpotmagfield}
\vec{A}(q_1,q_2)=(A_1;A_2;A_3)=\left(0;rB(q_2)\sin(q_1/r);0\right).
\end{equation}
With this choice of the gauge, the Hamiltonian reads \cite{Ferrari08}
\begin{equation}\label{eq:hamcyl}
\mathbb{H}=-\frac{\hbar ^2}{2m}\left(\frac{\partial^2}{\partial q_1^2}+\frac{\partial^2}{\partial q_2^2}\right)+\frac{\im e \hbar}{2m}\left(\frac{\partial}{\partial q_2}A_2+A_2 \frac{\partial}{\partial q_2}\right)+\frac{1}{2m}e^2 A_2^2-\frac{\hbar ^2}{8m r^2},
\end{equation}
where $e>0$ and $m$ are the electron elementary charge and mass, respectively.
The last term  in \eqz{\ref{eq:hamcyl}} is the potential arising from the (constant) curvature of the surface \cite{daCosta81} and
will be dropped in the following, since it amounts to a rigid energy shift.
Note that, for this particular surface and field configuration, the above equation can also be obtained by writing the Laplacian term contained in the Hamiltonian in cylindrical coordinates using the Peierls substitution \cite{Peierls33}, based on the principle of minimal coupling, ${\im\hbar}\frac{\partial}{\partial {\vec{q}}}\to({\im\hbar}\frac{\partial}{\partial {\vec{q}}}+e\vec{A})$.

\section{Homogeneous magnetic field\label{sec:constant}}
\begin{figure}
\centering
\includegraphics[width=7.5cm, angle=0]{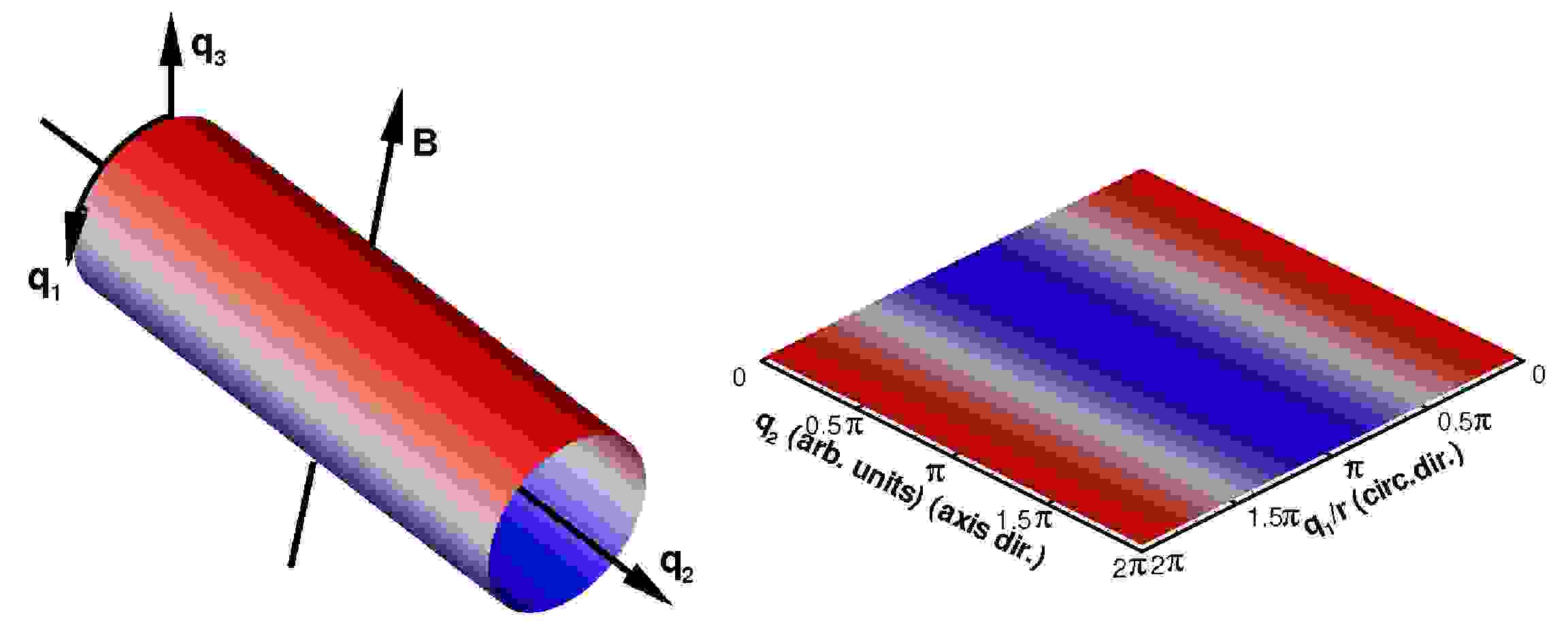}
\caption{\label{fig:homotube} The cylindrical surface $S$ in a homogeneous magnetic field $B$. The cylindrical reference frame used throughout the paper is indicated. The intensity of the field is shown in colour code both in space and as an open surface, with darker colours indicating a stronger field component normal to $S$, either entering (blue) or exiting (red) $S$.}
\end{figure}

\begin{figure}
\centering
\includegraphics[width=7.5cm, angle=0]{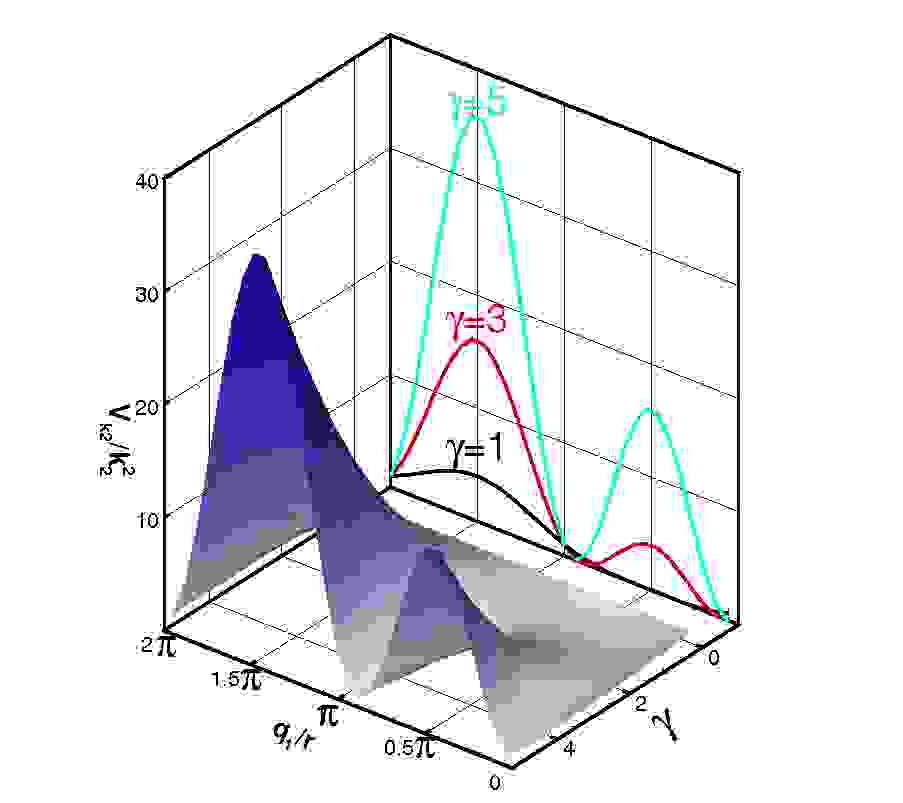}
\caption{\label{fig:xpot} The effective potential $V_{\ky}(q_1)/ \ky^2$, as a function of the position around the circumference, $q_1$ and of the coupling parameter $\gamma$. For low $\gamma$, $V_{\ky}(q_1)$ has one shallow minimum at $q_1=(\pi/2)r$ which bifurcates into two deeper minima
at larger values of $\gamma$. In the back of the graph, the potential is shown at three selected values of $\gamma$, as indicated.}
\end{figure}

We next apply the general formalism to the case of
a cylinder in a homogeneous magnetic field of intensity $B$.
Here, the vector potential in \eqz{\ref{eq:genvecpotmagfield}} reads
$
\vec{A}(q_1)=\left(0;rB\sin(q_1/r);0\right).
$
The geometry is shown in \Fig{\ref{fig:homotube}},
where the direction of the field is represented by the upward arrow, while the intensity of its component normal to $S$ is indicated by the light (low value) and dark (high value) colours.
The vector potential depends only on $q_1$, i.e. the position along the circumference of the cylinder, and does not break the translational invariance along $q_2$.
Since the field does not depend on $q_2$, the wave-function separates as
\begin{equation}\label{eq:seppsi}
\Psi(q_1,q_2) = \varphi(q_1) \ees^{\im \ky q_2},
\end{equation}
and the wave-vector $\ky$ along the axis is a good quantum number.
The Hamiltonian reads
\begin{equation}\label{eq:hamhom}
\mathbb{H} = \frac{\hbar^2}{2m}\left[-\frac{\partial^2}{\partial q_1^2} + V_{\ky}(q_1)\right],
\end{equation}
where
\begin{equation}\label{eq:effepot}
V_{\ky}(q_1) = \ky^2\left(1-\frac{eBr}{\hbar\ky}\sin{\frac{q_1}{r}}\right)^2
\end{equation}
is a wave-vector dependent 1D effective potential that wraps around the circumference of the cylinder.
Figure~\ref{fig:xpot} shows the profile of $V_{\ky}(q_1)/ \ky^2$ as a function of the dimensionless parameter
\begin{equation}
\gamma=\frac{eBr}{\hbar\ky},
\end{equation}
that gives the strength of the interaction between the charge and the field at a fixed wavevector value.
For $\gamma \lesssim 1$ the effective potential has one shallow minimum at $q_1=(\pi/2)r$, as a consequence the carriers tend to localise in a quasi-1D channel on one side of the cylinder, where the component of the field normal to $S$ is minimum, which side of the cylinder being decided by the relative sign of the wavevector and the field. For $\gamma \gtrsim 1$, $V_{\ky}(q_1)/ \ky^2$ has two minima, which for large $\gamma$ are located at $q_1=0$ and $q_1=\pi r$ (above and below the cylinder). In this regime, carriers form two quasi-1D channels, located where the component of the field normal to $S$ is maximum, and the field is either entering or exiting the surface.

\subsection{Energy levels\label{sec:enelevcost}}

For a given $\ky$,
we found the exact eigenstates of the Hamiltonian $\mathbb{H}$
by writing the $\varphi(q_1)$ component of $\Psi$
as linear combinations of the modes on the circumference
\begin{equation}\label{eq:unperturbed}
\varphi(q_1) = \frac{1}{\sqrt{2\pi}} \sum_n c_n \ees^{\im\frac{n}{r} q_1}.
\end{equation}
The zero-field energies of the 2D states are
\begin{equation}\label{eq:enunperturbed2}
\epsilon_n(\ky)=\frac{\hbar^2}{2m}\left(\frac{n^2}{r^2}+\ky^2\right),
\end{equation}
where, $n$ is an integer labelling the modes on the circumference, mixed by the magnetic field.
All reported calculations are obtained with $n=20$ in \eqz{\ref{eq:unperturbed}}.

In order to discuss the eigenstates of a carrier in the presence of a homogeneous field, it is convenient to define the dimensionless coupling parameter
\begin{equation}
\label{eq:alfac}
\alpha^{c}=r\sqrt{\frac{2\pi eB}{\hbar}},
\end{equation}
namely, the ratio between half of the circumference and the magnetic length, calculated averaging the intensity of the field over one half of the circumference. Clearly, this parameter describes the coupling between the field and the carrier.
Apart from the averaging, it corresponds to the one defined in Ref. \onlinecite{Ajiki93}. The energies of the eigenstates  scale with $\alpha^c$, i.e., the same energy is obtained for different values of the tube radius and field intensity as long as $r\sqrt{B}$ is constant.

\begin{figure}
\centering
\includegraphics[width=7.5cm, angle=0]{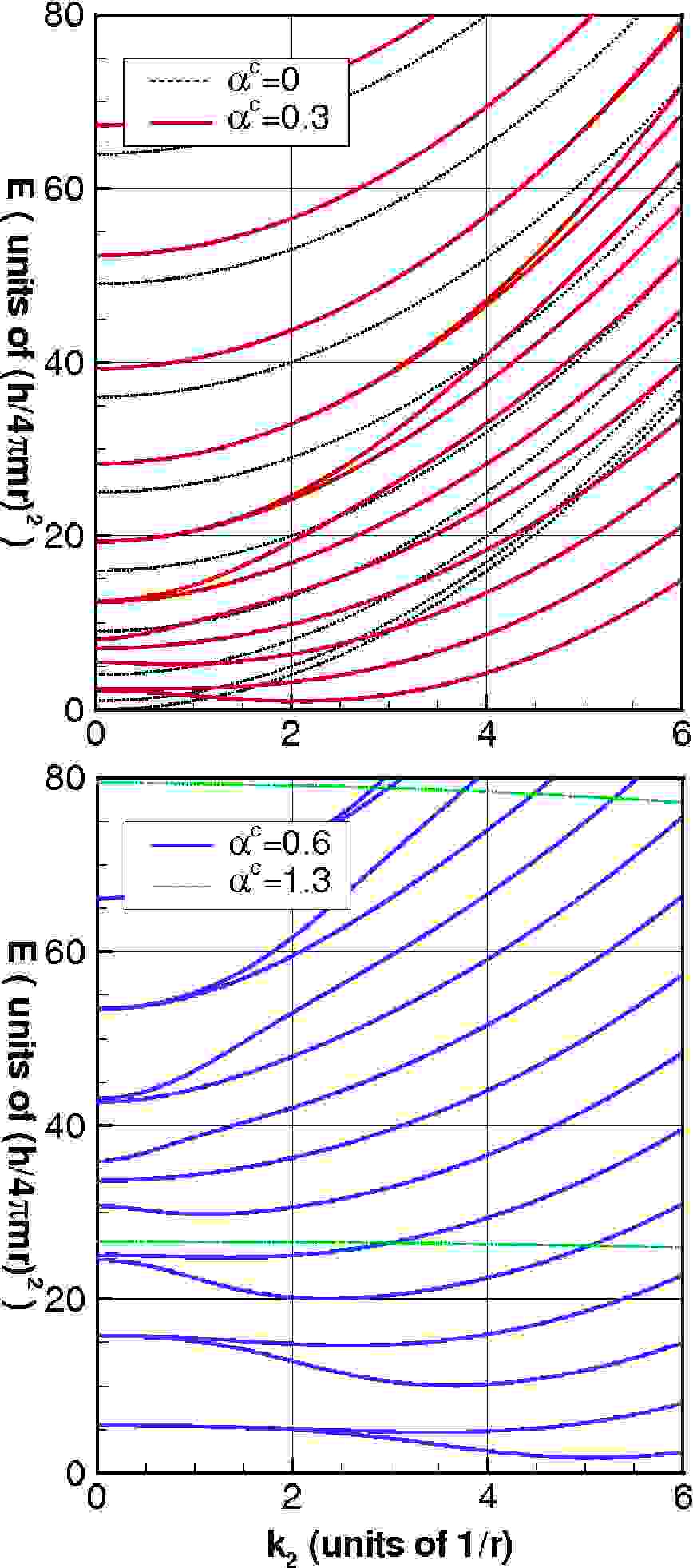}
\caption{\label{fig:bandshomo} Energy subbands at selected values of the coupling parameter $\alpha^{c}$. The parabolic subbands $\alpha^{c}=0$ (zero field) are doubly degenerate, except for the lowest one.}
\end{figure}

The subband structure for an homogeneous field is shown in \Fig{\ref{fig:bandshomo}}.
At $\alpha^c=0$ (zero field), the energy bands are given by \eqz{\ref{eq:enunperturbed2}}, therefore, all bands are doubly degenerate, except for the lowest one.
When a magnetic field is applied,
the double degeneracy is lifted by the orbital Zeeman splitting. For high values of $\alpha^{c}$, that is for sufficiently strong field at fixed tube radius, the subbands flatten at small $\ky$. By increasing $\alpha^{c}$, the energy becomes almost independent of the wave vector in a larger range, and approach the value of Landau levels in a planar 2D system. As we will see in the following
these states correspond to Landau-like states confined above and below the cylinder.

\subsection{Density of states and magnetic induced localisation\label{costdos}}

\begin{figure}
\centering
\includegraphics[width=7.5cm, angle=270]{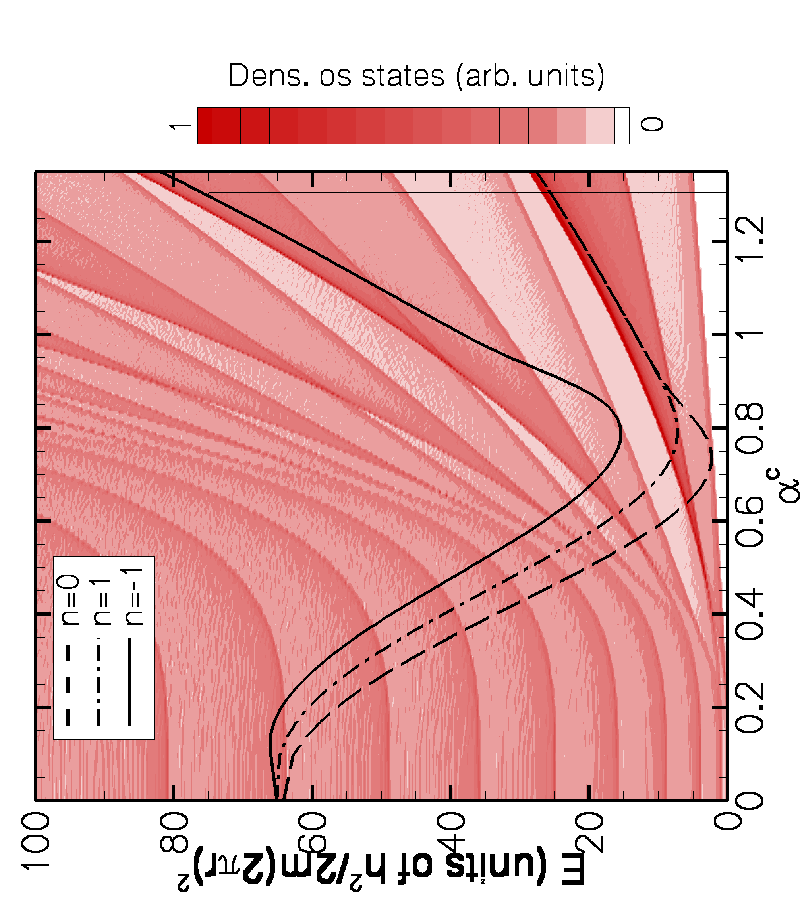}
\caption{\label{fig:const_dos} DOS as a function of the energy and the coupling parameter $\alpha^{c}$ (colour code given in the legend). For reference, we show three energy levels corresponding to $k_2 r = 8$, which evolve from the $n=0,\pm 1$ levels at $\alpha^c=0$, as indicated. 
The vertical line at $\alpha^{c}=1.3$ shows the position of the DOS reported in \Fig{\ref{fig:constmix}}.
}
\end{figure}

\begin{figure}
\centering
\includegraphics[width=7.5cm, angle=0]{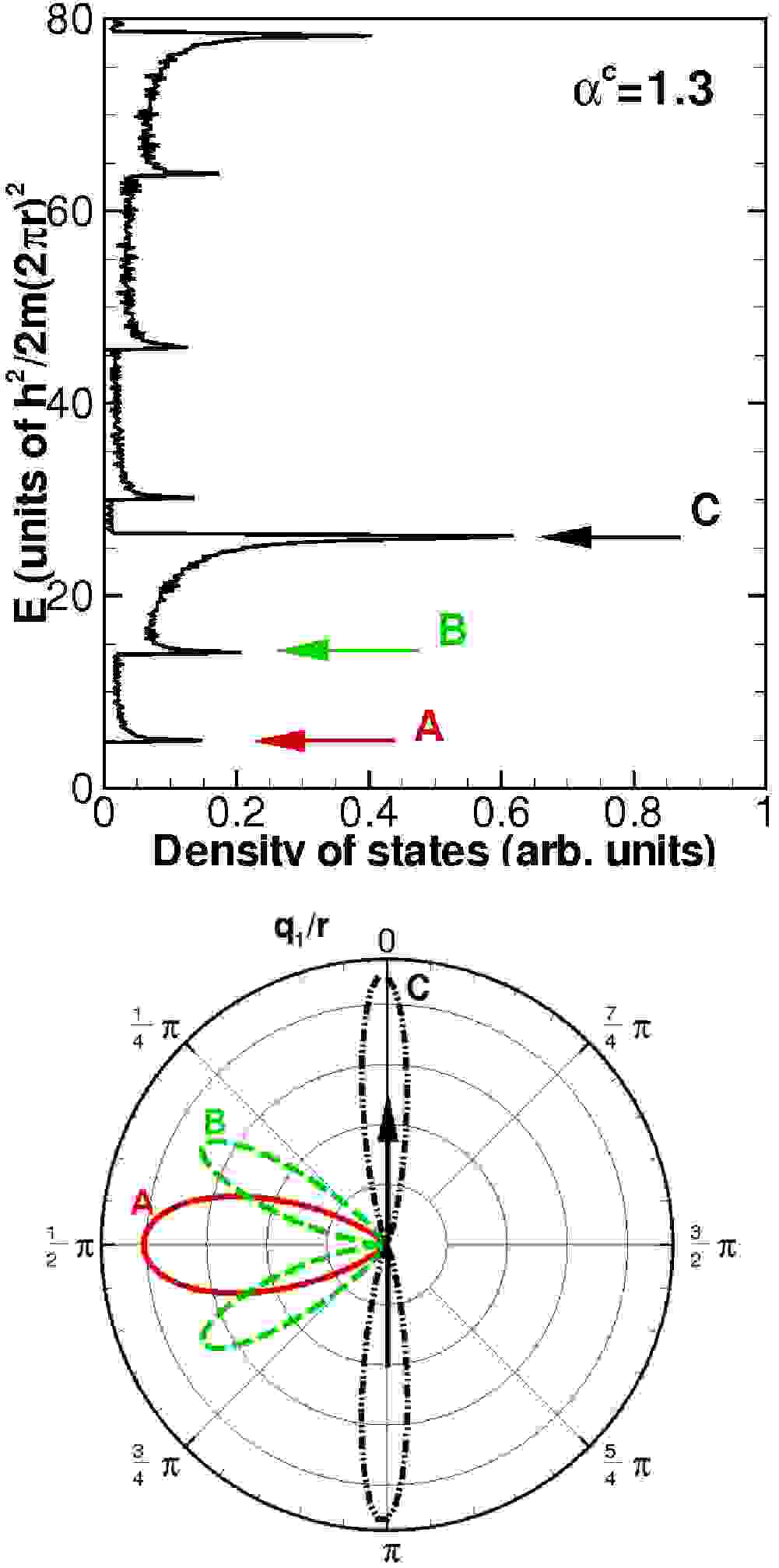}
\caption{\label{fig:constmix} Top: DOS at $\alpha^{c}=1.3$. The presence of many peaks indicating new 1D states are clearly visible. The states of peaks A and B correspond to states with $n=0$ and $n=1$ at zero field, respectively. The peak C contains states starting from both $n=0$ and $n=1$: for the given value of $\alpha^{c}$ they form 1D Landau levels.
Bottom: Probability densities of the states corresponding to the peaks in the top panel. The states A and B are 1D states localised in the region of the cylinder where $B$ is parallel to the surface: these are edge states driven in that position by Lorentz force. The state C belongs to a Landau level: these 1D states localise in the regions where $B$ is perpendicular to the surface. The direction of the field is indicated by the arrow.}
\end{figure}

The DOS, as a function of the energy and $\alpha^{c}$, is shown in \Fig{\ref{fig:const_dos}}.
We can recognise different regimes: i) at small $\alpha^c$ (low field), the orbital Zeeman splitting lifts the double degeneracy of the zero-field bands, ii) at intermediate fields, the energy of the states is lowered by the interaction with the field, and iii) at large fields, different subbands merge into highly degenerate levels, reminiscent of the corresponding 2D Landau level.
In  \Fig{\ref{fig:const_dos}}, the energy splitting due to the orbital Zeeman effect is shown for the two states $n=1$ (dash-dotted line) and $n=-1$ (solid line)
\endnote{
Here and in the following, each state $\varphi$ will be labelled with the
quantum number $n$ of the zero-field state from which it develops
continuously, as $B$ increases.  This unambiguous correspondence is
possible since the position of the zero-probability nodes of $\varphi$ is
not altered by the magnetic field.
}
, for a specific $\ky$ value. As the field is switched on, the degenerate levels split, then the energy decreases and reaches a minimum for a value of $\alpha^{c}$ that is larger for larger $n$ and $\ky$. The locus of the minima of all the states with the same $n$ (and different wavevectors) gives a peak in the DOS, indicating the formation of a 1D state.
In the top panel of \Fig{\ref{fig:constmix}} we show the DOS corresponding to $\alpha^{c}=1.3$. Peak A belongs to the locus of the minima of the $n=0$ states, and is strongly suggestive of the $\epsilon^{-1/2}$ behaviour of a quasi-1D system, with $\epsilon$ being the energy with respect to the band edge.
Analogously, peak B belongs to the locus of minima of the $n=1$ states. In the bottom panel of \Fig{\ref{fig:constmix}}, a polar plot shows the probability densities corresponding to peaks A, B and C (described later).
These 1D states localise in the region of the cylinder where the field is parallel to the surface. In fact, they are edge states, driven on one side of the cylinder by Lorentz force,  which side being determined by the sign of the charge and by the direction of $B$. A change in the relative sign of these parameters switches the localisation to the opposite side.
In this regime the potential $V (q_1)$ has one minimum, as shown in \Fig{\ref{fig:xpot}}.

For high values of $\alpha^{c}$, the energy levels gather into Landau levels, whose energy grows linearly with $B$. The first Landau level is formed, independently of $\ky$, by the states $n=0$ and $n=1$, the second one by the states $n=-1$ and $n=2$ and so on. The Landau levels are well recognisable in the density of states of \Fig{\ref{fig:const_dos}} where they appear as dark lines. In the top panel of \Fig{\ref{fig:constmix}}, the most prominent feature is peak C which corresponds to the formation of the first Landau level. Due to the finite curvature the energy levels acquire a finite dispersion, which gives rise to the 1D-like tail of the DOS on the low energy side, unlike the standard 2D Landau levels.
The probability density of a Landau state is localised in the regions where the magnetic field is perpendicular to $S$; as shown in the bottom panel of \Fig{\ref{fig:constmix}} it has two lobes, aligned with the magnetic field. In this case the region of localisation is independent of the charge sign and of the direction of the field: the 1D Landau states are always strips that run on the top and on the bottom of the cylinder.
However, since this effect is related to the formation of Landau levels on the cylinder, the strength of the localisation reaches a saturation for high $\alpha^c$. This means that the maximum localisation in the circumferential direction is independent of the cylinder radius, of the intensity of the magnetic field and of the wavevector $\ky$.

The detailed analysis given for the first three peaks of the DOS can be repeated for all other peaks. In general, as the value of $\alpha^{c}$ increases, each state
evolves in a 1D state of type A or B and then shrinks in a 1D Landau state.

\section{Spatially-modulated magnetic field\label{sec:variable}}

\begin{figure}
\centering
\includegraphics[width=7.5cm, angle=0]{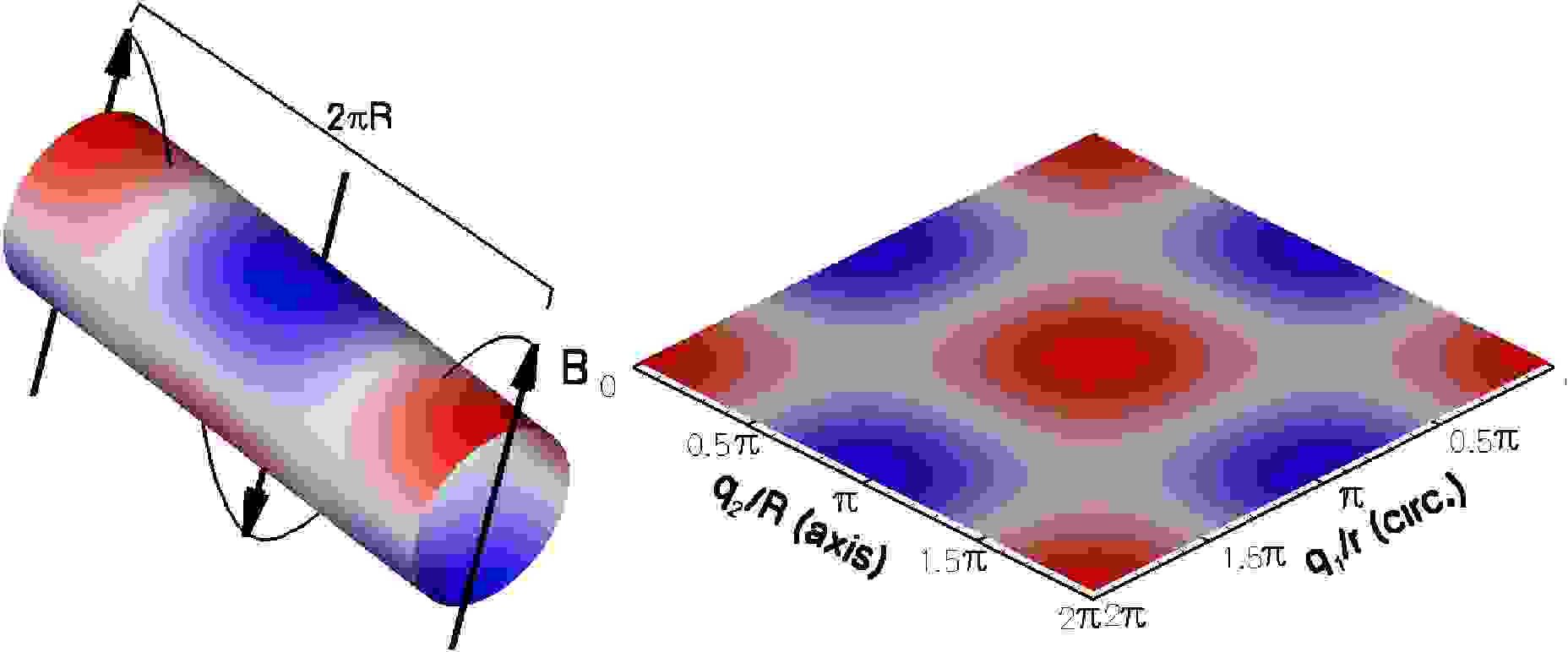}
\caption{\label{fig:varsys} The cylindrical surface $S$ of radius $r$ in a magnetic field periodically modulated with wavelength $2\pi R$ along the tube axis. The intensity of the field is shown in colour code both in 3D space and on the open surface, as in \Fig{\ref{fig:homotube}}. Lighter colours indicate regions where the vertical component of the field is zero, either because the field
is parallel to $S$ or because the intensity is zero.}
\end{figure}

We next consider the case of a magnetic field periodically modulated in intensity along the axis of the cylinder with zero average intensity.
To be specific, we consider a magnetic field whose intensity varies with a sinusoidal law along the axis. The vector potential in \eqz{\ref{eq:genvecpotmagfield}} reads
\begin{equation}
\label{eq:varvecpotmagfield}
\vec{A}(q_1,q_2)=(0;rB\cos(q_2/R)\sin(q_1/r);0),
\end{equation}
where $2\pi R$ is the wavelength of the spatial modulation of the field along $q_2$ and $B$ its maximum intensity.
In \Fig{\ref{fig:varsys}}, arrows indicate the direction of the modulated magnetic field, while the intensity of its component perpendicular to $S$ is shown in colour code. Note the square pattern formed by the white regions, where the perpendicular component of the field vanishes, either because the field is parallel to $S$ or because it has a vanishing intensity there.

Since the vector potential now depends on both $q_1$ and $q_2$, the wavefunction can not be separated as in \eqz{\ref{eq:seppsi}}, and we have to solve a fully 2D problem.
However, since the Hamiltonian is periodic in $q_2$, we can apply the Bloch's theorem to a periodic system with a  1D unit cell of length $2\pi R$ along $q_2$.
The exact eigenstates $\chi(q_1,q_2)$ are found as a linear combination of the zero field states, as
\begin{equation}\label{eq:bloch}
\chi(q_1,q_2)=\frac{1}{2\pi}\sum_{nm} c_{nm}(\ky)\ees^{\im\frac{n}{r}q_1}\ees^{\im(\ky+mG)q_2},
\end{equation}
where $G=1/R$ and $-G/2 \le \ky \le G/2$. The corresponding zero-field energies are
\begin{equation}
\epsilon_{nm}(\ky)=\frac{\hbar^2}{2m}\left(\frac{n^2}{r^2}+\frac{m^2}{R^2}+\ky^2\right).
\end{equation}
Here, the quantum numbers $n$ and $m$ are integers representing the mode numbers around the circumference and along $q_2$ in the 1D unit cell, respectively. When a finite field is applied, the $n$ and $m$ modes are mixed. In these section, the results are obtained by taking $n=m=30$ in \eqz{\ref{eq:bloch}}.

One useful parameter to characterise the system is $\alpha^{v}=\Phi / \Phi_0$, that is the ratio between $\Phi$, namely the magnetic flux through a cell $(\pi r\times\pi R)$,
and the magnetic flux quantum $\Phi_0=2\pi\hbar/e$.
Note that the field is not constant within the cell $(\pi r\times\pi R)$.
Therefore, we have
\begin{equation}
\Phi=\frac{1}{\pi^2 r R}\int_{0}^{\pi r}\int_{0}^{\pi R} B(q_1,q_2)\de q_1 \de q_2 = 4BRr,
\end{equation}
and
\begin{equation}
\alpha^v=\frac{2eBRr}{\pi\hbar}.
\end{equation}
The parameter $\alpha^v$ gives the strength of the coupling between the field and the carriers, and plays a role analogous to $\alpha^c$ in the homogeneous case of the previous section. The coupling increases linearly with the intensity of $B$ and with the spatial periodicities $r$ and $R$.
Indeed, $\alpha^v$ is defined in analogy to other works involving magnetic field applied to a 2D system \cite{Hofstadter76,Olariu85,Aoki92,Kim92,Ajiki94}. In Ref. \onlinecite{Hofstadter76}, $\alpha^{v}$ is alternatively interpreted as
the ratio between the period of motion of an electron with crystal momentum $\hbar/\sqrt{rR}$ (which is $2\pi rRm/\hbar$) and the reciprocal of the average cyclotron frequency $4eB/\pi^2 m$.

In general terms, the vector potential creates an effective 2D potential on the surface $S$ that tends to localise the wavefunction where the component of the field orthogonal to $S$ is zero (lighter regions in \Fig{\ref{fig:varsys}}). Specifically, the low energy states of the carriers will be mainly localised at the intersections of the above stripes. These quasi-0D regions are connected by tunnelling in a 2D network.
The ratio $\rho=r/R$ between the cylinder radius $r$ and the length of the field modulation $R$ identifies different regimes, whether $\rho<1$, $\rho\approx 1$ or $\rho>1$, as we show next.

\subsection{Ring-like localisation\label{sec:var_noose}}

\begin{figure*}
\centering
\includegraphics[width=12.cm, angle=0]{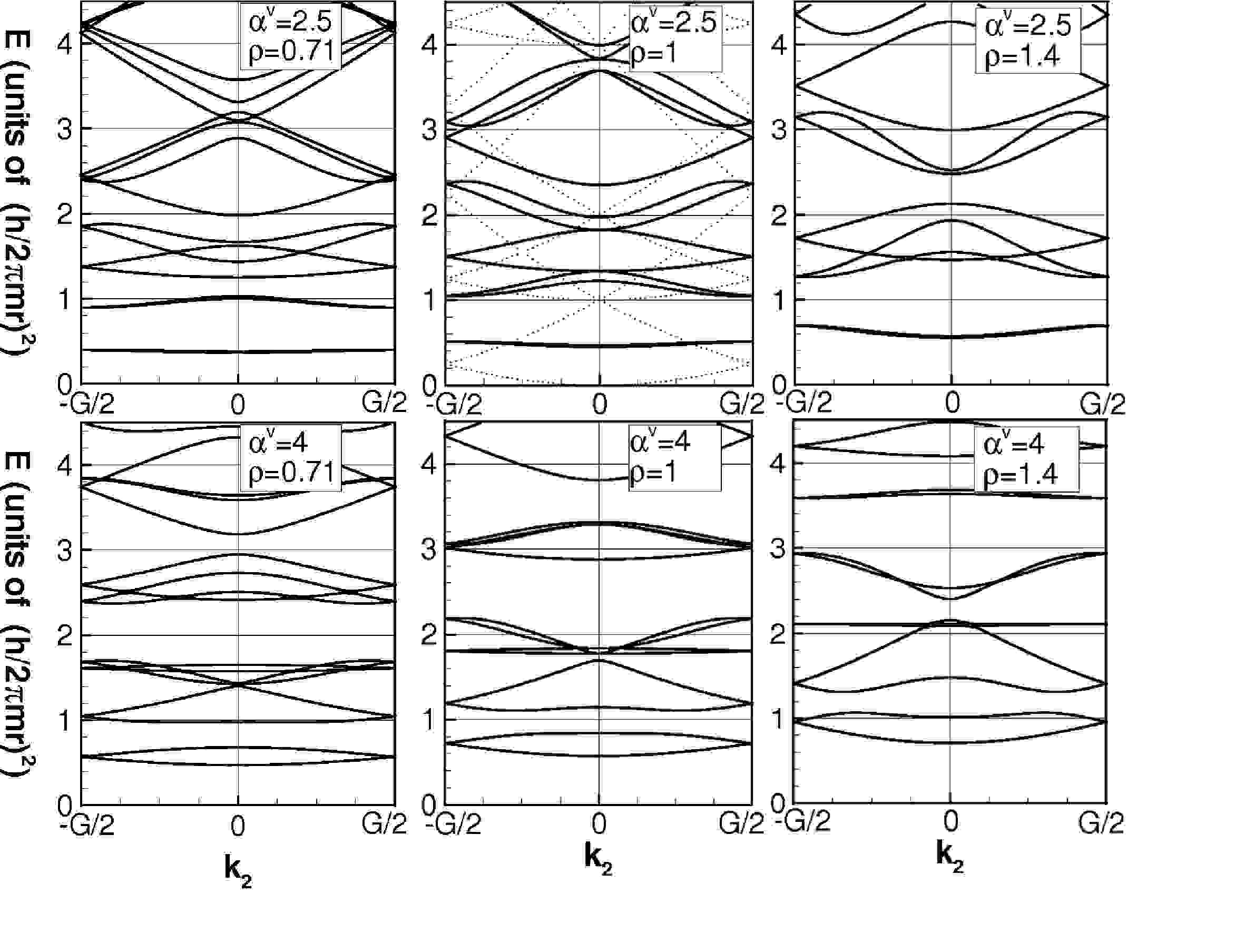}
\caption{\label{fig:bands_mod} Energy subbands for the three selected values of $\rho = 0.71, 1, 1.4$ (left, centre, right column), each at $\alpha^{v}=2.5, 4$ (upper,  lower row).
In the upper centre panel the parabolic subbands at zero field are also shown with dotted lines, for comparison.
}
\end{figure*}

Let us start from the case $\rho<1$, namely the field is slowly modulated with respect to the cylinder diameter, and $r<R$. All the results of this subsection are for $\rho=0.71$. 
The left panels of \Fig{\ref{fig:bands_mod}} show the energy bands  at two values of $\alpha^v$. Compared to the gapless parabolic band structure at zero magnetic field, reported in the central upper panel (dotted lines), the finite-field band structure is characterised by large gaps in the low energy range. Furthermore, the lowest subbands are almost flat for the values of $\alpha^{v}$ shown here, which indicates carrier localisation.

\begin{figure}
\centering
\includegraphics[width=7.5cm, angle=270]{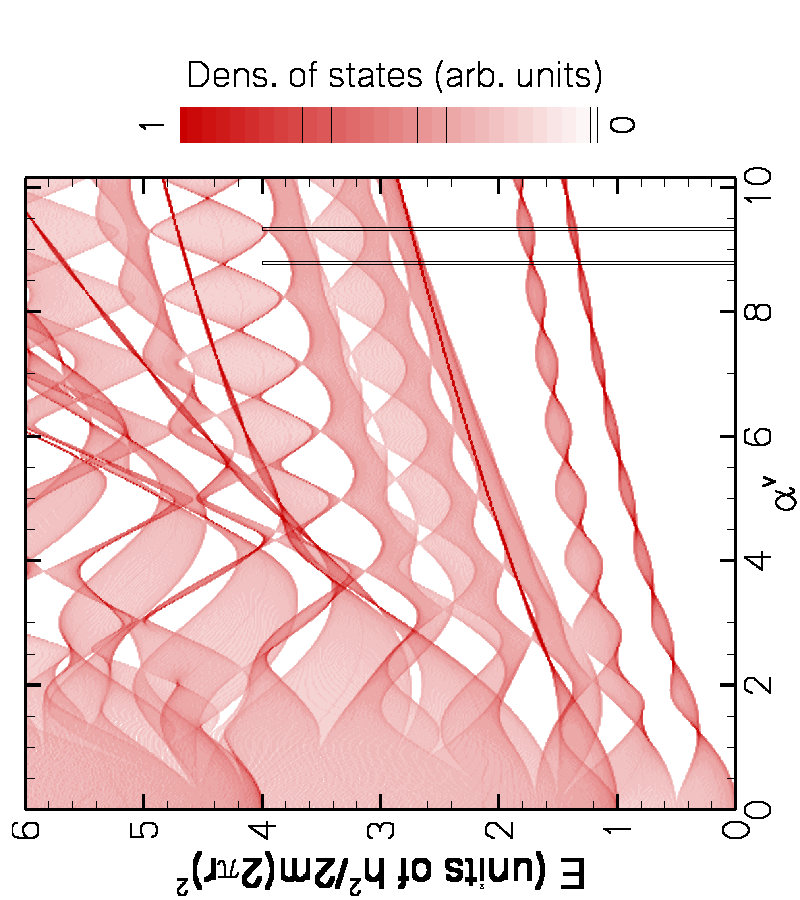}
\caption{\label{fig:dens_noose} DOS at $\rho=0.71$ (see Sec.~\ref{sec:var_noose}) as a function of the energy and the coupling parameter $\alpha^{v}$ (colour code given in the legend). The DOS shown in \Fig{\ref{fig:anellimix}}
correspond to the two vertical lines at $\alpha^{v}=8.77$ and $9.33$.
}
\end{figure}

\begin{figure}
\centering
\includegraphics[width=7.5cm, angle=0]{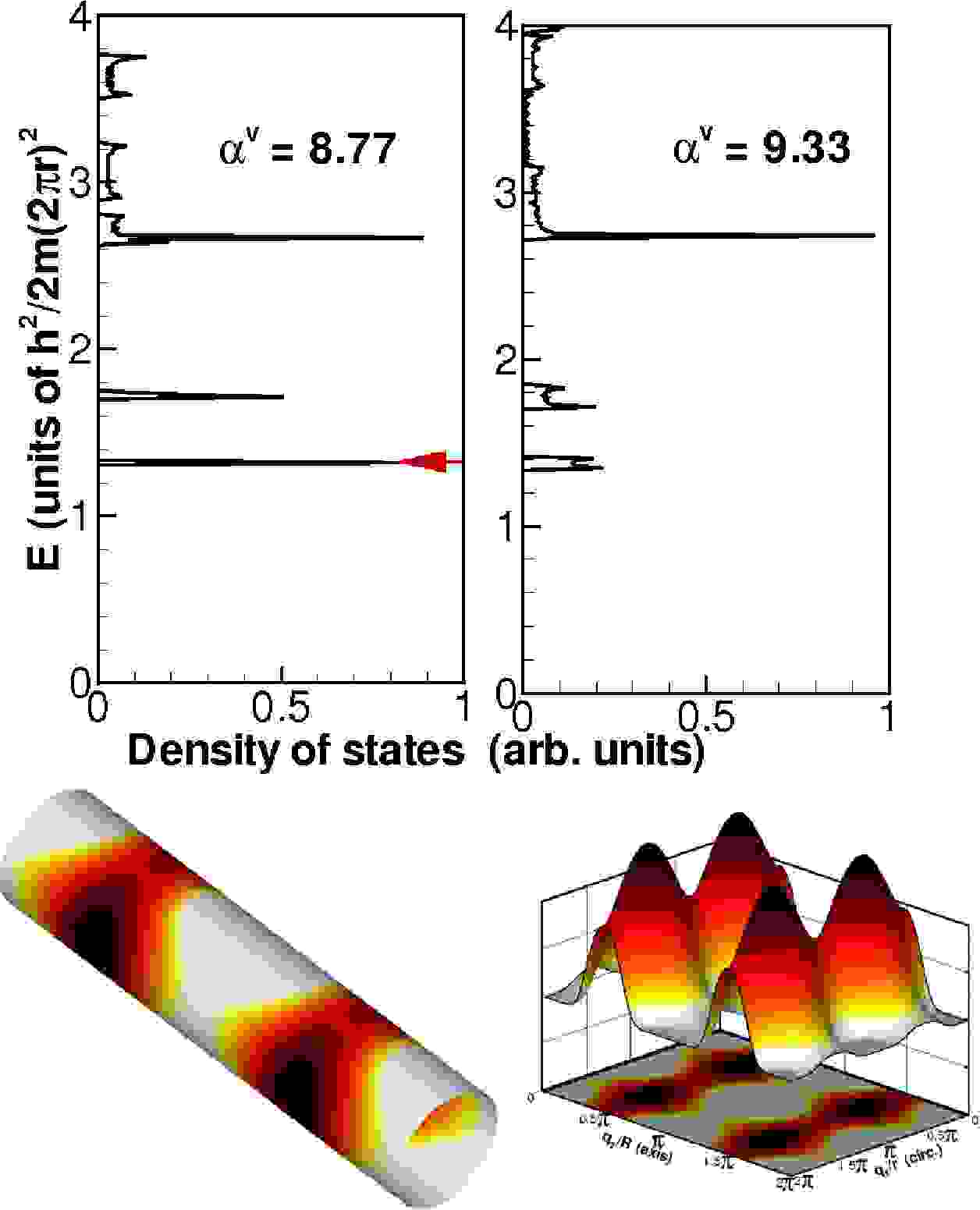}
\caption{\label{fig:anellimix} Top: DOS at $\rho=0.71$ for $\alpha^{v}=8.77$ and $9.33$. 
Bottom: Probability density for a state belonging to the peak of indicated with the arrow in the top panel. Darker colour means higher density.
The applied field rearranges the C2DEG in a lattice of quasi-0D states, connected in a ring-like shape around the tube, and weakly coupled along the tube.}
\end{figure}

In \Fig{\ref{fig:dens_noose}} we show  the DOS  as a function of the coupling parameter $\alpha^{v}$. This clearly shows the opening of many energy gaps with an amplitude which strongly depend on $\alpha^{v}$. At somewhat regular values, the two lowest subbands are completely flat. The modulation in the DOS and the occurrence of peaks (darker lines in \Fig{\ref{fig:dens_noose}}) is also shown in the top panel of \Fig{\ref{fig:anellimix}} for two specific values of $\alpha^{v}$.

In the bottom panel of \Fig{\ref{fig:anellimix}} we show the probability density on the tube for a state belonging to the peak in the top panel indicated with the arrow.
The charge density is mainly distributed in a superlattice of quasi-0D states induced by the magnetic field, localised at the intersections of the regions where the perpendicular component
of $B$ is zero (see \Fig{\ref{fig:varsys}}). The localisation is stronger for increasing $\alpha^v$. For the present case of $\rho<1$,
the asymmetric shape of the unit cell $(\pi r\times\pi R)$ turns into an asymmetric tunnelling between the quasi-0D regions, where charge is mainly localised. At this value of $\alpha^{v}$, tunnelling is almost suppressed along the axis direction, but it is present around the circumference. The effect is that the 2DEG is rearranged in an arrays of weakly coupled pairs of quantum dots.
Also, note that the positions of the dots along the axis is independent of the sign of the charges, and, since the confinement occurs in the zeros of the effective field, the position is also independent of the direction of the field.

\subsection{Strip-like localisation\label{sec:var_strips}}

Next we analyse the case $\rho>1$, namely, the field modulation is rapid with respect to the tube diameter, and $R>r$. 
All numerical results presented in this subsection are for $\rho=1.40$. The energy subbands are shown in \Fig{\ref{fig:bands_mod}} (right panels) for two values of $\alpha^v$.
Again, the magnetic field affects the dispersion, and opens energy gaps. Almost flat bands can be observed for specific values of the $\alpha^{v}$ parameter and localised states will be observed on the surface of the cylinder.

\begin{figure}
\centering
\includegraphics[width=7.5cm, angle=270]{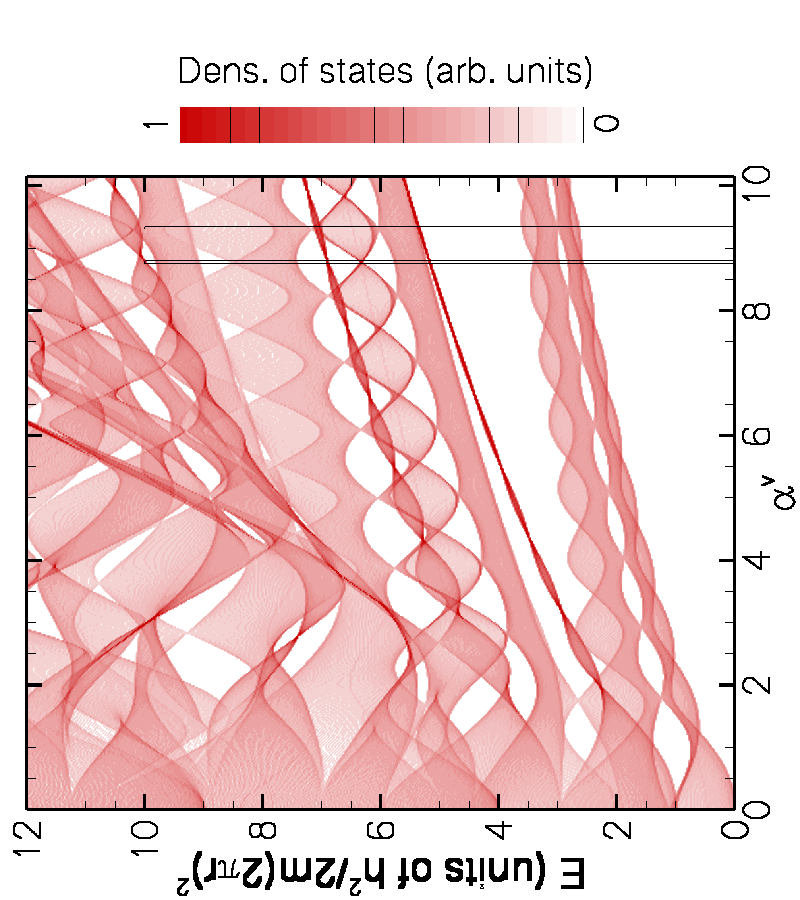}
\caption{\label{fig:dens_strip} DOS at $\rho=1.40$ (see Sec.~\ref{sec:var_strips}) as a function of the energy and the coupling parameter $\alpha^{v}$ (colour code given in the legend). The DOS shown in \Fig{\ref{fig:striscemix}}
correspond to the two vertical lines at $\alpha^{v}=8.77$ and $9.33$.
}
\end{figure}

\begin{figure}
\centering
\includegraphics[width=7.5cm, angle=0]{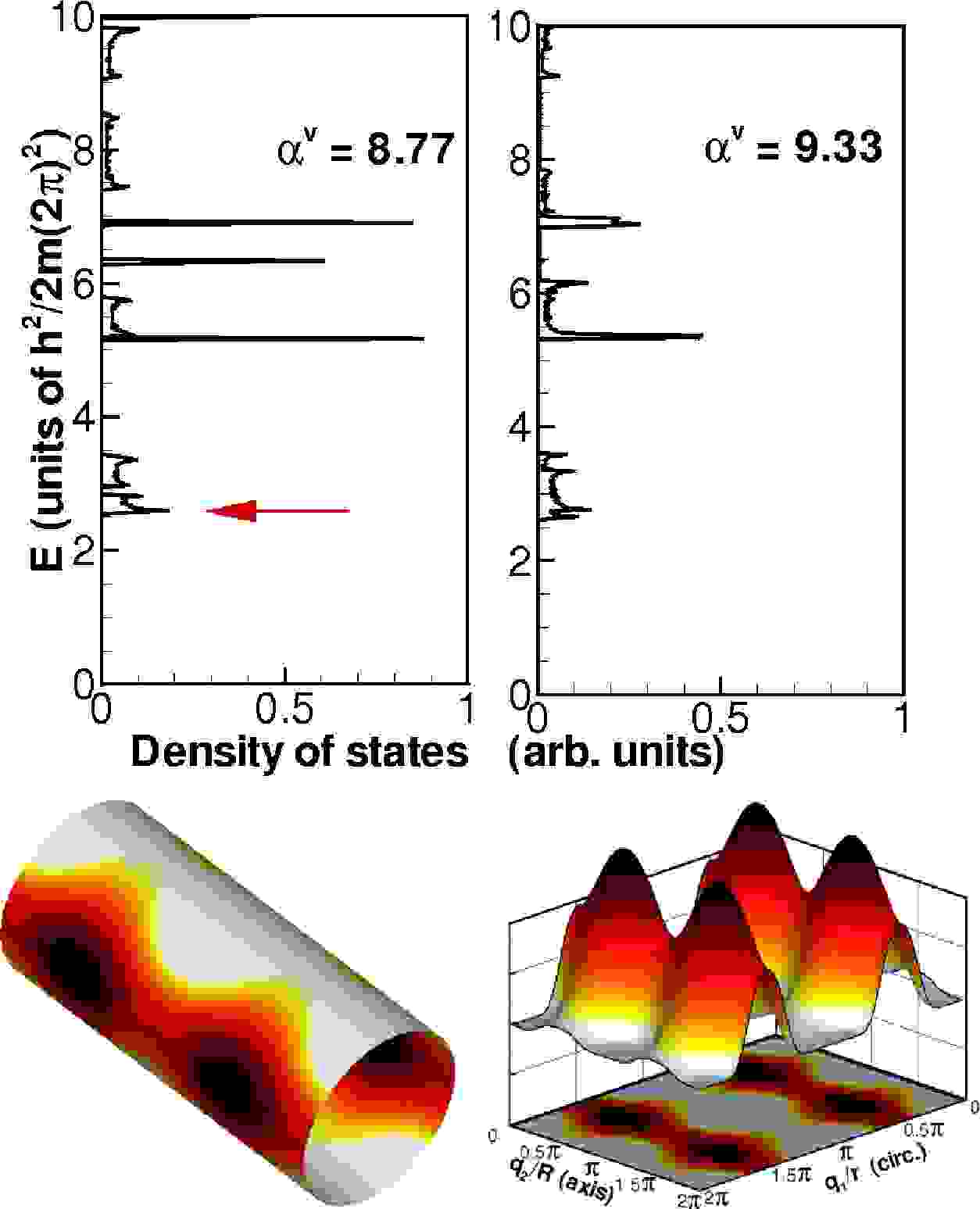}
\caption{\label{fig:striscemix} Top: DOS at $\rho=1.40$ for $\alpha^{v}=8.77$ and $9.33$.
Bottom: Probability density for a state belonging to the peak indicated with the arrow in the top panel. Darker colour means higher density.
In this case, tunnelling between the dots is not negligible along the axis direction, while it is almost suppressed around the circumference.}
\end{figure}

The DOS as a function of the parameter $\alpha^v$ is shown in \Fig{\ref{fig:dens_strip}} and in the top panel of \Fig{\ref{fig:striscemix}}. Again, the magnetic field has the main effect of opening gaps not present at zero field, although the gap pattern is different from \Fig{\ref{fig:dens_noose}}.

In the bottom panel of \Fig{\ref{fig:striscemix}}, we show the probability density on the tube for a state belonging to the peak indicated with the arrow in the top panel. 
A rearrangement of the charge density in a lattice of quasi-0D regions is obtained, analogously to the $\rho<1$ case. However, the unit cell $(\pi r\times\pi R)$ is now shorter along the cylinder axis, and tunnelling between the quasi-0D states is larger along the axis direction. 
Therefore, the magnetic field induces a localisation of the carriers in two arrays of tunnel-coupled dots along the axis direction, the two arrays on opposite sides of the cylinder being weakly coupled by tunnelling. Also in this case, the localisation is stronger for increasing $\alpha^v$ and does not depend on the sign of the charge nor on the direction of the field.

\subsection{Dot-like localisation\label{sec:var_rings}}

The $\rho=1$ case is an admittedly difficult condition to be obtained exactly, but it is discussed here for completeness and as an example of the intermediate regime $r\approx R$.
The energy bands are shown in the central panels of \Fig{\ref{fig:bands_mod}}. Again, for $\alpha^{v}=2.5$ a flat band is present, and the width of this band is modulated by the magnetic field which affects tunnelling. 

\begin{figure}
\centering
\includegraphics[width=7.5cm, angle=270]{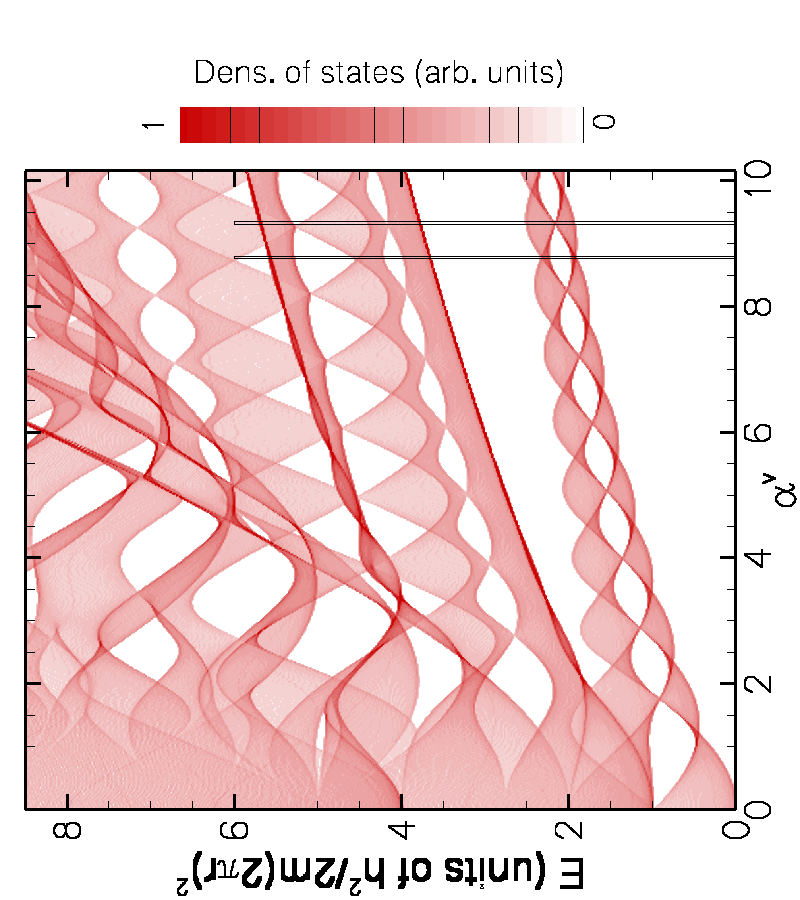}
\caption{\label{fig:dens_rings} DOS at $\rho=1$ (see Sec.~\ref{sec:var_rings}) as a function of the energy and the coupling parameter $\alpha^{v}$ (colour code given in the legend). The DOS shown in \Fig{\ref{fig:quadratomix}}
correspond to the two vertical lines at $\alpha^{v}=8.77$ and $9.33$.
}
\end{figure}

\begin{figure}
\centering
\includegraphics[width=7.5cm, angle=0]{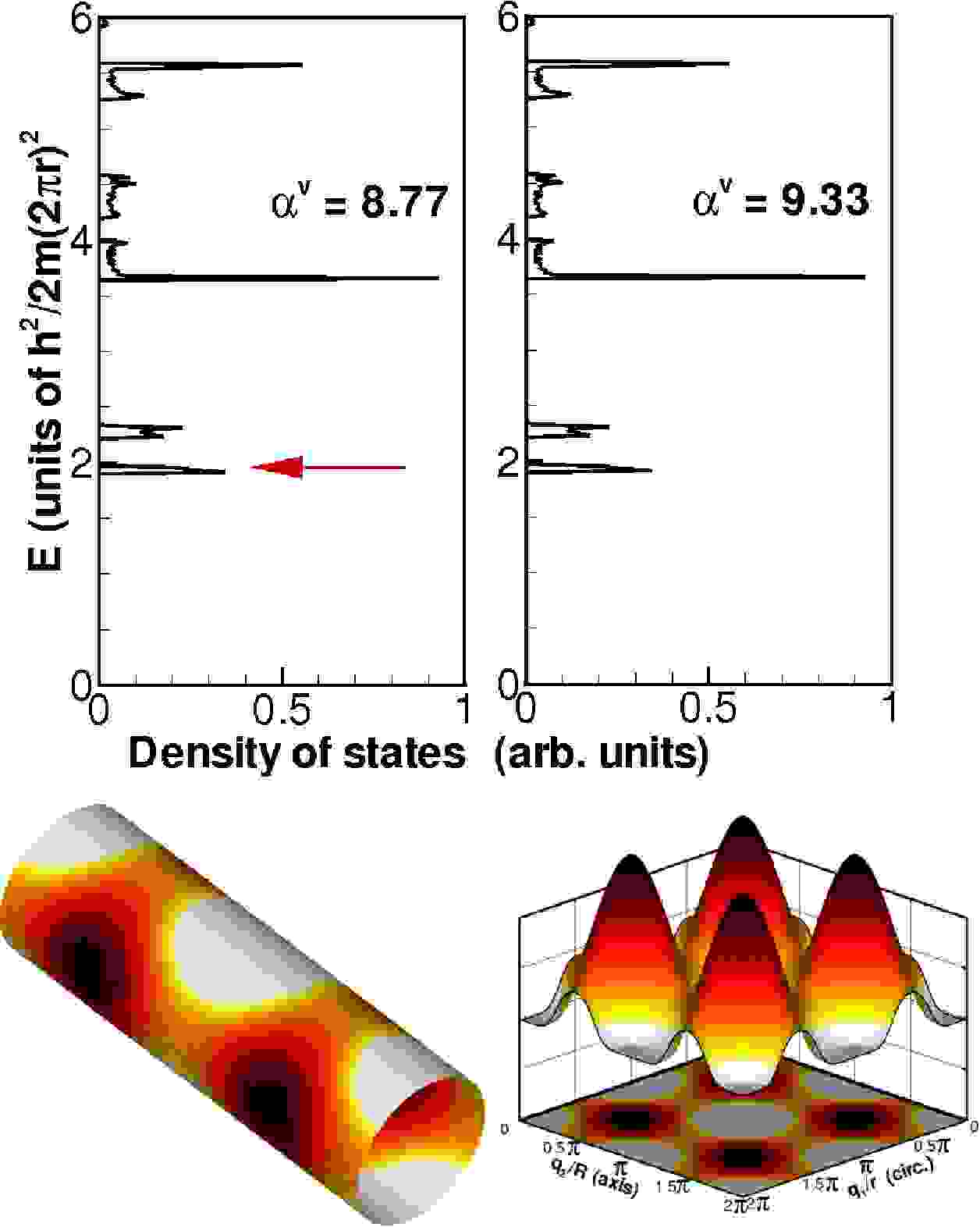}
\caption{\label{fig:quadratomix} Top: DOS at $\rho=1$ for $\alpha^{v}=8.77$ and $9.33$.
Bottom:  Probability density for a state belonging to the peak indicated with the arrow in the top panel. Darker colour means higher density.}
\end{figure}

The DOS is shown in \Fig{\ref{fig:dens_rings}} as a function of the parameter $\alpha^v$.
Now, the modulation of the DOS shows regular patterns of energy gaps and peaks, periodic with $\alpha^{v}$. A detailed analysis of these features will be given in section \ref{sec:lan_aha}. The DOS at selected values of $\alpha^{v}$ is also shown in \Fig{\ref{fig:quadratomix}}, where the peaks and gaps are clearly visible.  

In the bottom panel of \Fig{\ref{fig:quadratomix}} we show the probability density on the tube for a state belonging to the peak indicated with the arrow in the top panel.
Due to the additional symmetry of the $(\pi r \times \pi R)$ unit cell, the tunnelling probability between the quasi-0D states is now the same along the axis and around the circumference directions, and a superlattice of dots each connected by four arms to neighbouring dots is formed.

To summarise, we expect that at $r\approx R$ a periodically modulated magnetic field creates a square lattice of 0D states, the localisation being stronger for larger intensities, and independent from the direction and the sign of the carriers.
When $r$ substantially deviates from $R$, the tunnel coupling between the dots increases either along cylinder axis ($r>R$), creating two weakly coupled 1D arrays of quantum dots, or along the cylinder circumference ($r<R$), making the system more similar to a 1D array of weakly coupled quantum rings.

\subsection{Insights from the energy landscape\label{sec:lan_aha}}

Having established in the previous sections how the energy levels, and the ensuing DOS are strongly affected  by the intensity and modulation of the magnetic fields, we now look more closely to the overall behaviour with respect to the interaction parameter $\alpha^v$. This will give us a deeper insight into the physics governing this system and exposes the analogies and consistencies with other physical situations.

\subsubsection{Aharonov-Bohm oscillations}

We first focus on the low-energy part of the spectrum, and consider for definiteness the case $\rho=1$.
A peculiar characteristic of the DOS (\Fig{\ref{fig:dens_rings}}) is the oscillation of the energy levels so to form a plait. Figure~\ref{fig:varlandau} shows the energy levels at $\ky=0$, with the oscillatory behaviour of the lowest levels with $\alpha^v$.
This trend is characteristic of many electronic systems under the effect of a magnetic field, and it is a typical fingerprint of Aharonov-Bohm type behaviour\cite{Olariu85,Fuhrer01}. In this specific device,
the dots and the arms connecting them are regions of zero field which constitute a loop around a region of non-zero perpendicular component of the magnetic field, as shown in the bottom panel of \Fig{\ref{fig:quadratomix}}.
Clearly, this explains qualitatively the shift of the energy levels with $\alpha^v$, although our explicit calculation must take into account that the rings are not well defined for low values of $\alpha^v$, since the tunnelling between the dots can be very high, and the dots themselves are quite large. Furthermore, the direction and intensity of the field is not constant throughout the ring and, finally, the shape of the ring is not round but more square-like.
This explains why the minimum of the energy of the ground state is not constant against $\alpha^v$, as it would be in a text-book Aharonov-Bohm system. Furthermore the periodicity of the oscillations is not strictly an integer of the ratio $\Phi/\Phi_0$.
Note that in this system the Aharonov-Bohm ring is not physically defined in the absence of the field, and is induced by the same magnetic field and its interplay with the geometry of the C2DEG.

\subsubsection{Landau levels}

\begin{figure*}
\centering
\includegraphics[width=7.5cm, angle=0]{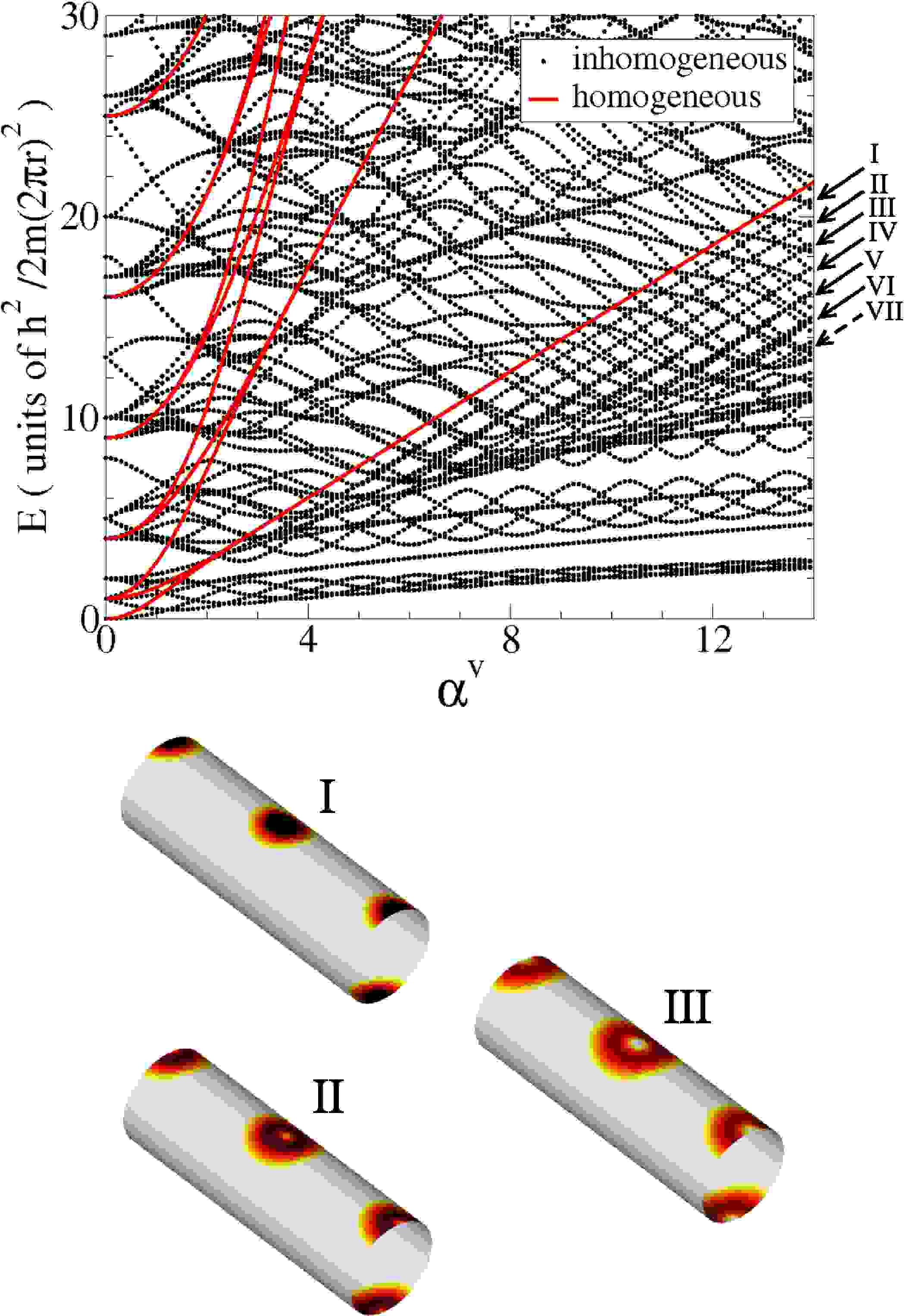}
\caption{\label{fig:varlandau} Top: the energy levels at $\ky=0$ are presented as a function of $\alpha^v$, in the case of $\rho=1$, up to the 100$^{th}$ level.
The bands corresponding to the Landau levels are indicated by the arrows on the right axis. The Landau levels for the corresponding case with an homogeneous field are shown with a solid line.
Bottom: the probability densities of the system for the Landau levels I, II and III (see the arrows on the left plot) are shown. Darker colour means higher density. The carriers are confined in the regions with the field perpendicular to the cylinder surface.}
\end{figure*}

In the Sec.~\ref{sec:constant} we have connected the energy levels at high homogeneous magnetic field with the formation of Landau levels on the surface of the tube.
In the inhomogeneous case the lowest energy states are confined close to the regions where the field is parallel to the surface or its intensity goes to zero (we recall that the average field is zero in the present investigation). Since the Landau levels cannot appear in regions of zero perpendicular field, they are not involved in the formation of low energy states, but they form at higher energy.
Figure~\ref{fig:varlandau} compares the $\ky =0$ energy levels of the homogeneous and inhomogeneous cases, for the same $r$, in a broad energy range (up to the 100$^{th}$ level).
A set of levels, indicated by arrows and roman numerals, show a linear shift with $\alpha^{v}$ with the same slope as the first Landau level of the homogeneous case. Plots of the carrier densities show that these states correspond to  carriers which are confined in the region where the field is perpendicular to the cylinder surface, that is the regions circulated by the Aharonov-Bohm-like rings which localise the low energy states. These are highly degenerate states, due to the flat dispersion with respect to $k_y$ and $n$. Clearly, the degeneracy is not the same as for genuine 2D Landau levels, due to the periodicity which is imposed by the modulated field and the cylindrical symmetry. The degeneracy can also be traced to the charge density (\Fig{\ref{fig:varlandau}}, right panels) which is redistributed in dots which are well separated, tunnelling between these regions being completely suppressed.

\section{Conclusions\label{conclusions}}

A magnetic field, either homogeneous or periodically modulated affects the dimensionality of carrier states in a C2DEG, depending on the ratio between the diameter of the nanostructure, the magnetic length and, possibly, the wavelength of the field modulation.
In the case of an homogeneous field, the extended states of the cylinder are redistributed in 1D channels by a magnetic field: two regimes are identified, with different localisation of the 1D channels, corresponding to different $\ky$ and field intensity. In the case of a field periodically modulated in space, localisation into a superlattice of tunnel-coupled quasi-0D states is induced. The dots may be connected on the surface $S$ to form a ring-like, or a strip-like shape, this being determined by the ratio between the radius of the cylinder and the wavelength of the field modulation. Dimensionless parameters in terms of field intensity, periodicity and cylinder diameter identify the different regimes. In particular, the interplay between the field intensity and the periodicity, and cylinder diameter leads to a strong rearrangement of the energy band structure, which would result in peculiar transport properties of the system.
We also reconciled the energy band structure of the C2DEG to the familiar case of Landau level formation in planar 2DEGs.
In the homogeneous field case the formation of Landau levels on the tube surface, is responsible of the confinement in quasi-1D channels on opposite sides of the cylinder; in the modulated field case, the Landau levels are still present, with energies much higher than the ground state, while the low energy spectrum is characterised by an Aharonov-Bohm behaviour.

\begin{acknowledgments}
The authors are pleased to thank Giampalo Cuoghi for the fruitful discussions.

This work has been partially supported by by project FIRB-RBIN04EY74.
\end{acknowledgments}

\bibliography{ferrari}

\end{document}